\documentclass{PoS}
\usepackage{bm,amsmath,amssymb}
\usepackage{graphicx}
\usepackage{subfigure}
\usepackage{color}
\usepackage{soul}
\setulcolor{blue}
\setstcolor{red}
\sethlcolor{yellow}
\newcommand \etc {{\sl etc.}}

\newcommand \ie {{\sl i.e.}}

\newcommand \fig[1] {Fig.\ \ref{#1}}

\newcommand \eq[1] {Eq.\ (\ref{#1})}
\newcommand \beq {\begin{equation}}
\newcommand \eeq {\end{equation}}
\newcommand \beqa {\begin{eqnarray}}
\newcommand \eeqa {\end{eqnarray}}
\newcommand \lrnd {\left(}
\newcommand \rrnd {\right)}
\newcommand \lsq {\left[}
\newcommand \rsq {\right]}

\newcommand \lang {\left\langle}
\newcommand \rang {\right\rangle}

\newcommand \nn {\nonumber}
\newcommand \hmu {\hat{\mu}}
\title{Deconfinement of strangeness and freeze-out from charge fluctuations}
\ShortTitle{Deconfinement and freeze-out}

\author{\speaker{Swagato Mukherjee}
        \thanks{This work has been supported in part through contract
        DE-AC02-98CH10886 with the U.S. Department of Energy.} \\
        Physics Department, Brookhaven National Laboratory, Upton, NY 11973, USA\\
        E-mail: \email{swagato@bnl.gov}
        }

\author{Mathias Wagner\\
        Fakult\"at f\"ur Physik, Universit\"at Bielefeld, D-33615 Bielefeld, Germany\\
        E-mail: \email{mwagner@physik.uni-bielefeld.de}
        }

\abstract{We use Lattice QCD calculations of fluctuations and correlations of various
conserved charges to show that the deconfinement of strangeness takes place in the
chiral crossover region of QCD; however, inside the quark-gluon plasma strange quarks
remain strongly interacting at least up to temperatures twice the QCD crossover
temperature.  Further, we discuss how the freeze-out parameters of heavy-ion
collisions can be determined in a model-independent way through direct comparisons
between experimentally measured higher order cumulants of conserved charges and
corresponding Lattice QCD calculations. Utilizing the preliminary data from the STAR
and PHENIX experiments we illustrate this method. Although, the Lattice QCD based
determinations of the freeze-out parameters utilizing data sets of different
experiments and different observables are currently not consistent with each other,
it is tantalizing to see that all the observed freeze-out parameters lie very close
to the chiral/deconfinement crossover region of QCD.}
          
\FullConference{8th International Workshop on Critical Point and Onset of
Deconfinement,\\ March 11 to 15, 2013\\ Napa, California, USA}

\begin{document}
%------------------------------------------------------------------------------------
%       introduction
%------------------------------------------------------------------------------------
\section{Introduction}

Understanding the nature of strongly interacting matter demands a detailed knowledge
regarding the phase structure of Quantum ChromoDynamics (QCD), the underlying theory
of strong interaction.  Several experimental programs, such as the recent Beam Energy
Scan (BES) program at the Relativistic Heavy Ion Collider (RHIC), Brookhaven National
Laboratory as well as future experiments at the upcoming FAIR and NICA facilities,
have been dedicated to uncover the phase diagram of QCD under extreme conditions,
i.e., high temperatures and/or large densities. On the other hand, to complete our
knowledge of the QCD phase diagram it is also necessary to supplement these
experimental endeavors with first-principle based theoretical calculations. Over the
years Lattice QCD (LQCD) has emerged as the most successful technique for performing
non-perturbative, parameter free theoretical calculations starting from the QCD
Lagrangian.

In this talk we discuss two recent LQCD calculations that closely complement the
experimental explorations of the QCD phase diagram. First, we present evidence that
at zero baryon density the deconfinement of strangeness takes place in conjunction
with the chiral crossover. Next, we describe a method for a model independent
determination of the freeze-out temperature and chemical potentials of heavy ion
collision experiments through a direct comparisons between the state-of-the-art LQCD
calculations and the experimentally measured cumulants of charge fluctuations.

To address these issues we rely on the LQCD computations of the generalized
susceptibilities of the conserved charges
\beq
\chi_{mn}^{XY} = \left. \frac{\partial^{(m+n)} [p(\hmu_X,\hmu_Y)/T^4]} 
{\partial \hmu_X^m \partial \hmu_Y^n} \right|_{\vec{\mu}=0}
\ ,
\label{eq:susc}
\eeq
where $\vec{\mu}=(\mu_B,\mu_S,\mu_Q)$ are respectively the baryon number, strangeness
and electric charge chemical potentials and $X,Y=B,S,Q$. For brevity, we use the
notations $\chi_{0n}^{XY}\equiv\chi_n^Y$ and $\chi_{m0}^{XY}\equiv\chi_m^X$.  These
generalized susceptibilities are related to the cumulants, such as the mean ($M_X$),
variance ($\sigma_X$), skewness ($S_X$) and kurtosis ($\kappa_X$), of the
fluctuations of the conserved charge. For example--- $VT^3\chi_1^Q=\lang
N_Q\rang=M_Q$, $VT^3\chi_2^Q=\lang(\delta N_Q)^2\rang=\sigma_Q^2$,
$VT^3\chi_3^Q=\lang(\delta N_Q)^3\rang=\sigma_Q^3S_Q$ and $VT^3\chi_4^Q=\lang(\delta
N_Q)^4\rang - 3\lang(\delta N_Q)^2\rang^2=\sigma_Q^4\kappa_Q$; $V$ being the volume,
$T$ the temperature and $N_X$ the net charge with $\delta N_X = N_X -\lang N_X\rang$.
Details of the LQCD calculations presented here can be found in
\cite{strange,freeze,hrg,Tc}. 

%------------------------------------------------------------------------------------
%       strangeness
%------------------------------------------------------------------------------------
\section{Deconfinement of strangeness and strange degrees of freedom inside quark
gluon plasma}

\begin{figure}[t!]
\subfigure[]{ \label{fig:v12}
\includegraphics[height=0.25\textheight,width=0.48\textwidth]{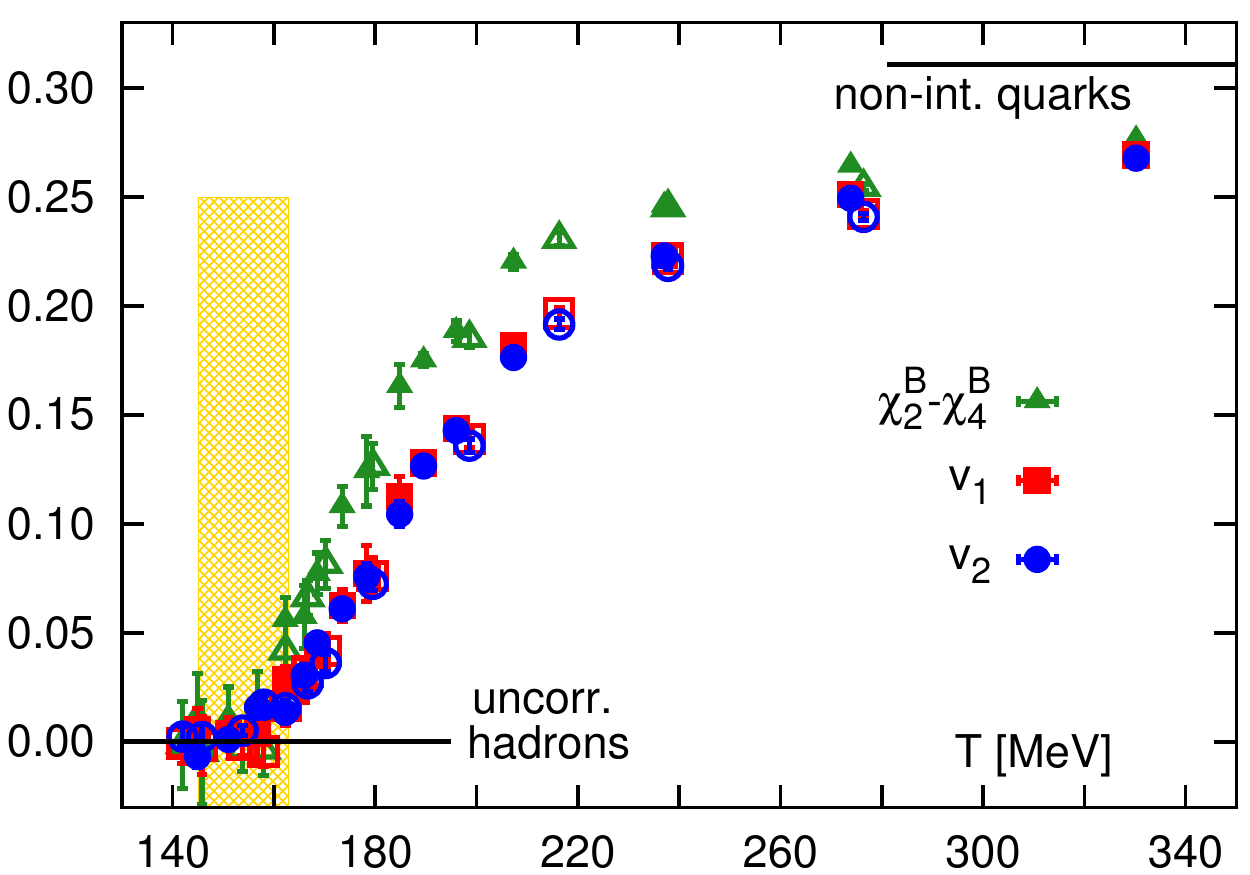} }
\subfigure[]{ \label{fig:BS_QS}
\includegraphics[height=0.25\textheight,width=0.48\textwidth]{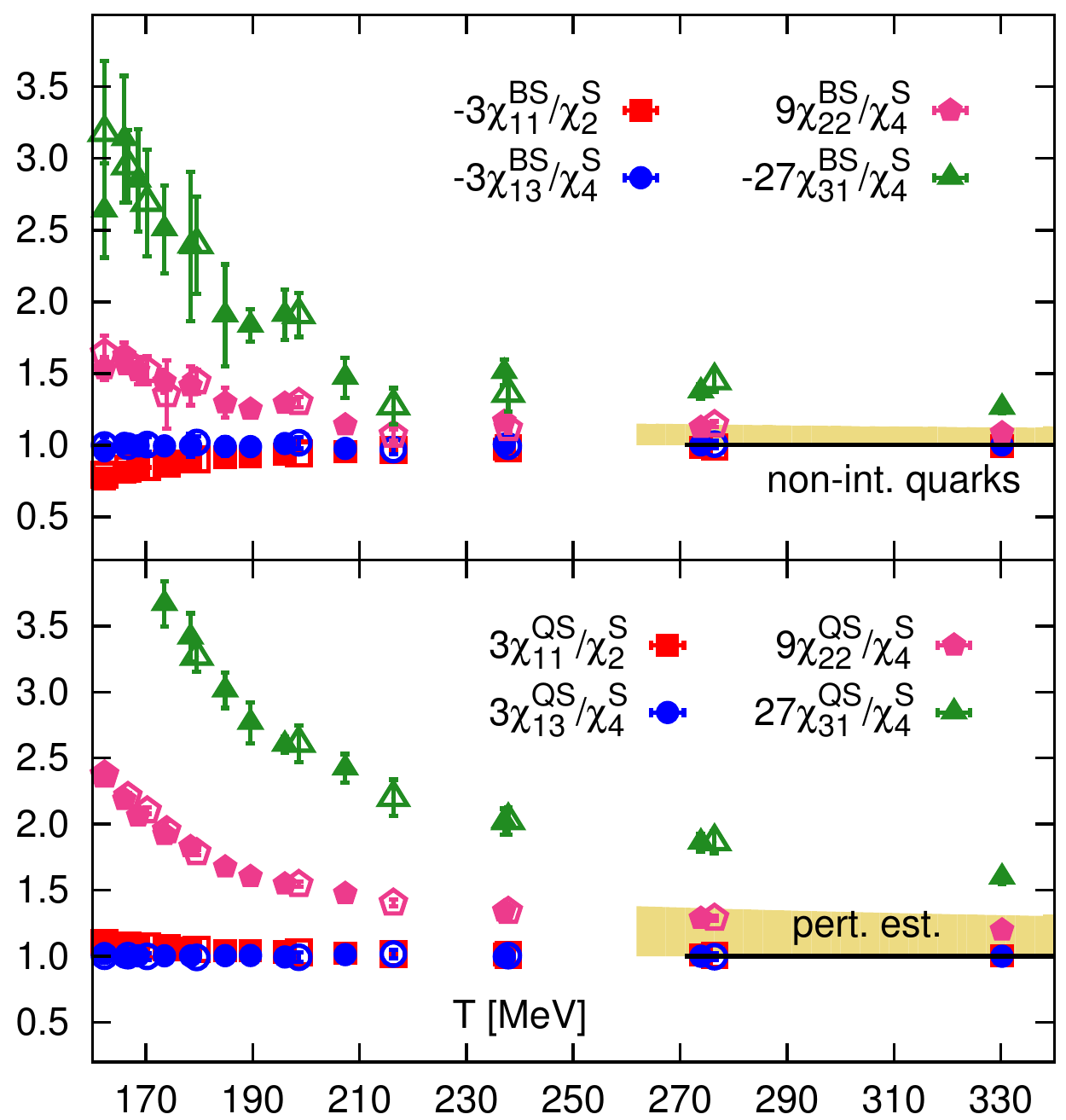} }
\caption{ \emph{(a)} Two combinations, $v_1$ and $v_2$, of strangeness fluctuations
and baryon-strangeness correlations that vanish identically if the sDoF are described
by an uncorrelated gas of hadrons.  The solid lines at low and high temperatures
indicate the two limiting scenarios when the dof are described by an uncorrelated
hadron gas and non-interacting massless quark gas, respectively. The chiral crossover
temperature $T_c=154(9)$ MeV \cite{Tc} is indicated by the shaded region. Further
shown is the difference of quadratic and quartic baryon number fluctuations,
$\chi_2^B-\chi_4^B$, that also vanishes when the baryon number carrying degrees of
freedom are strange and non-strange baryons.  \emph{(b)}  Baryon-strangeness (top)
and electric charge-strangeness correlations (bottom), normalized by the strangeness
fluctuations and scaled by appropriate powers of the baryonic and electric charges of
a strange quark such that in a non-interacting massless quark gas all these
observables are unity (indicated by the lines at high temperatures). The shaded
regions indicate the range of perturbative estimates for all these observables
obtained using one-loop re-summed Hard Thermal Loop calculations \cite{Vuorinen}. The
open and filled symbols are the LQCD results \cite{strange} for two different lattice
spacings corresponding to the temporal extents $N_\tau=6$ and $8$, respectively.}
\end{figure}

In an uncorrelated gas of hadrons, such as in the Hadron Resonance Gas (HRG) model
\cite{pbm}, the dimensionless partial pressure, $P_{S}\equiv(p-p_{S=0})/T^4$, of the
strangeness carrying Degrees of Freedom (sDoF)  can be written as 
\beqa
P^{HRG}_S(\hmu_B,\hmu_S) &=& P^{HRG}_{|S|=1,M} \cosh(\hmu_S) 
+ P^{HRG}_{|S|=1,B} \cosh(\hmu_B-\hmu_S)
\nn \\
&+& P^{HRG}_{|S|=2,B} \cosh(\hmu_B-2\hmu_S)
+ P^{HRG}_{|S|=3,B} \cosh(\hmu_B-3\hmu_S) 
\;,
\label{eq:P-HRG}
\eeqa
where $P^{HRG}_{|S|=1,M}$ is the partial pressure of all $|S|=1$ mesons and
$P^{HRG}_{|S|=i,B}$ are the partial pressures of all $|S|=i$ ($i=1,2,3$) baryons at
$\vec{\mu}=0$.  In the above expression the (classical) Boltzmann approximation has
been employed for all strange hadrons as their masses are substantially larger than
the temperature range of interest. To probe whether the sDoF are associated with
integral strangeness and baryon number, as in the case of a hadron gas, we introduce
\cite{strange} the following combinations consisting of the strangeness fluctuations
and baryon-strangeness correlations
\beq
v_1 = \chi_{31}^{BS} - \chi_{11}^{BS} 
\;, \qquad \mathrm{and} \qquad 
v_2 = \frac{1}{3} \lsq \chi_2^S - \chi_4^S \rsq - \lsq 2 \chi_{13}^{BS} - 
4 \chi_{22}^{BS} - 2 \chi_{31}^{BS} \rsq
\;. \label{eq:v12}
\eeq
From \eq{eq:susc} and \eq{eq:P-HRG} it is easy to see that these two combinations
vanish exactly for an uncorrelated gas of hadrons, \ie\ $v_1^{HRG}=v_2^{HRG}=0$.
Since in a hadron gas the sDoF are associated with $|B|=1$, baryon-strangeness
correlations differing by even numbers of $\mu_B$ derivatives are identical, leading
to $v_1^{HRG}=0$. On the other hand, the two parenthetically enclosed combinations in
the expression of $v_2$ individually amount to the partial pressure of the $|S|=2,3$
baryons, giving $v_2^{HRG}=0$. The LQCD results \cite{strange} for these two
combinations are shown in \fig{fig:v12}. There we also draw the difference between
the quadratic ($\chi_2^B$) and the quartic ($\chi_4^B$) baryon number fluctuations
that also receive contributions from the light up and down quarks. This combination
also vanishes when the strange and light quarks are are confined within hadrons
following the same argument as for $v_1^{HRG}$. It is clear that the LQCD data for
$v_1$, $v_2$ and $\chi_2^B-\chi_4^B$ are consistent with zero up to chiral crossover
temperature  $T_c=154(9)$ MeV \cite{Tc} and show a rapid increase towards their
non-interacting massless quark gas values above the $T_c$ region. The sDoF behave
quite similarly as those involving the light quarks; they are consistent with a
hadronic description up to $T_c$ and show rapid departures above $T_c$. The vanishing
values of these observables at low temperatures do not depend on the mass spectrum of
the relevant degrees of freedom, as long as they are uncorrelated and the Boltzmann
approximation is applicable. It stems from the fact that they carry integer
strangeness $|S|=0,1,2,3$ and baryon number $|B|=0,1$. Thus, altogether, LQCD
provides strong indications that up to the chiral crossover strangeness remains
confined within hadrons and the deconfinement of strangeness takes place around the
chiral crossover temperatures.

Based on experimental results from the RHIC and LHC by now it has been generally
accepted that for moderately high temperatures the deconfined  Quark Gluon Plasma
(QGP) phase of QCD remains strongly interacting. It is an intriguing question whether
such a strongly interacting QGP consists of quasi-quarks or its is strongly coupled
system devoid of a quasi-particle description. To elucidate the nature of sDoF inside
the QGP at moderately high temperatures we study the correlations of net strangeness
fluctuations with fluctuations of net baryon number and electric charge.  For
weakly/non-interacting quasi-quarks strangeness $S=-1$ always comes with a baryon
number of $B=1/3$ and an electric charge of $Q=-1/3$. Thus,
\beq
\frac{\chi_{mn}^{BS}}{\chi_{m+n}^S} = \frac{(-1)^n}{3^m}
\;,\quad\mathrm{and}\quad
\frac{\chi_{mn}^{QS}}{\chi_{m+n}^S} = \frac{(-1)^{m+n}}{3^m}
\;,\qquad \mathrm{where}\quad m,n>0,\ m+n=2,4 \;.
\label{eq:BS-QS}
\eeq
LQCD results \cite{strange} for these ratios, scaled by the proper powers of the
fractional baryonic and electric charges, are shown in \fig{fig:BS_QS}. Each of these
scaled baryon/charge-strangeness correlations should be unity for a massless gas of
non-interacting quasi-quarks. For $T_c\lesssim T\lesssim2T_c$  the LQCD results for
the second order baryon/charge-strangeness correlations are far from the values
expected for non-interacting quarks. To illustrate the effects of weak interactions
among the quasi-quarks we also indicate (shaded regions at high temperatures) the
ranges of values for these ratios as predicted for the weakly interacting
quasi-quarks. These values have been calculated from the re-summed Hard Thermal Loop
perturbation theory at the one-loop order \cite{Vuorinen}, using one-loop running
coupling obtained at the scales between $\pi T$ and $4\pi T$.  LQCD results involving
correlations of strangeness with higher power of baryon number and electric charge
clearly indicate that a description in terms of weakly interacting quasi-quarks
cannot be valid for temperatures $T\lesssim2T_c$. Thus LQCD results provide
unambiguous evidence that sDoF inside QGP can only become compatible with the
weakly/non-interacting quasi-quarks only for temperatures $T\gtrsim2T_c$.

%------------------------------------------------------------------------------------
%       freeze-out
%------------------------------------------------------------------------------------
\section{LQCD based model independent determination of freeze-out conditions in
heavy ion collisions}

\begin{figure}[t!]
\subfigure[]{ \label{fig:qi} 
\includegraphics[height=0.25\textheight,width=0.31\textwidth]{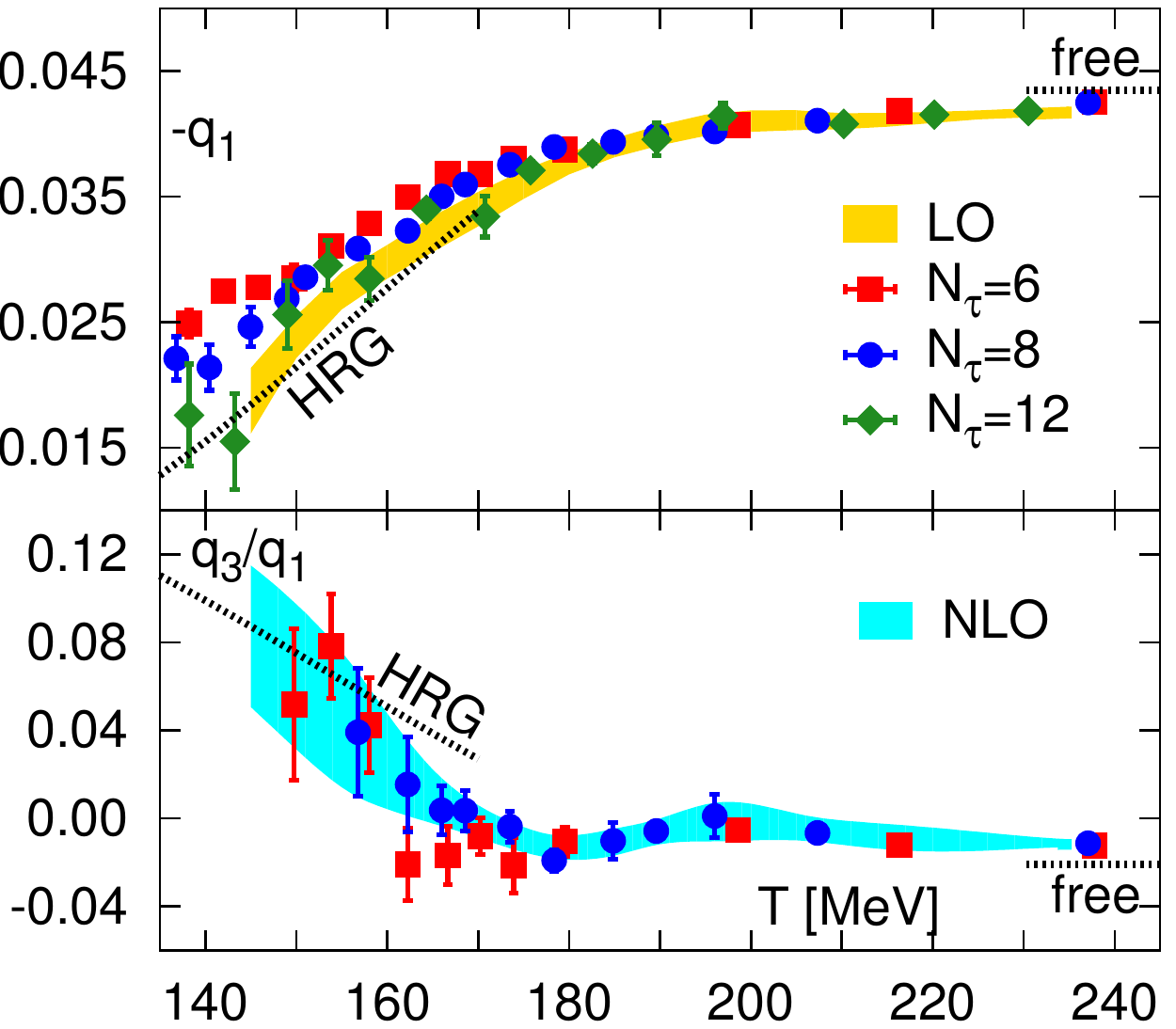} }
\subfigure[]{ \label{fig:si} 
\includegraphics[height=0.25\textheight,width=0.31\textwidth]{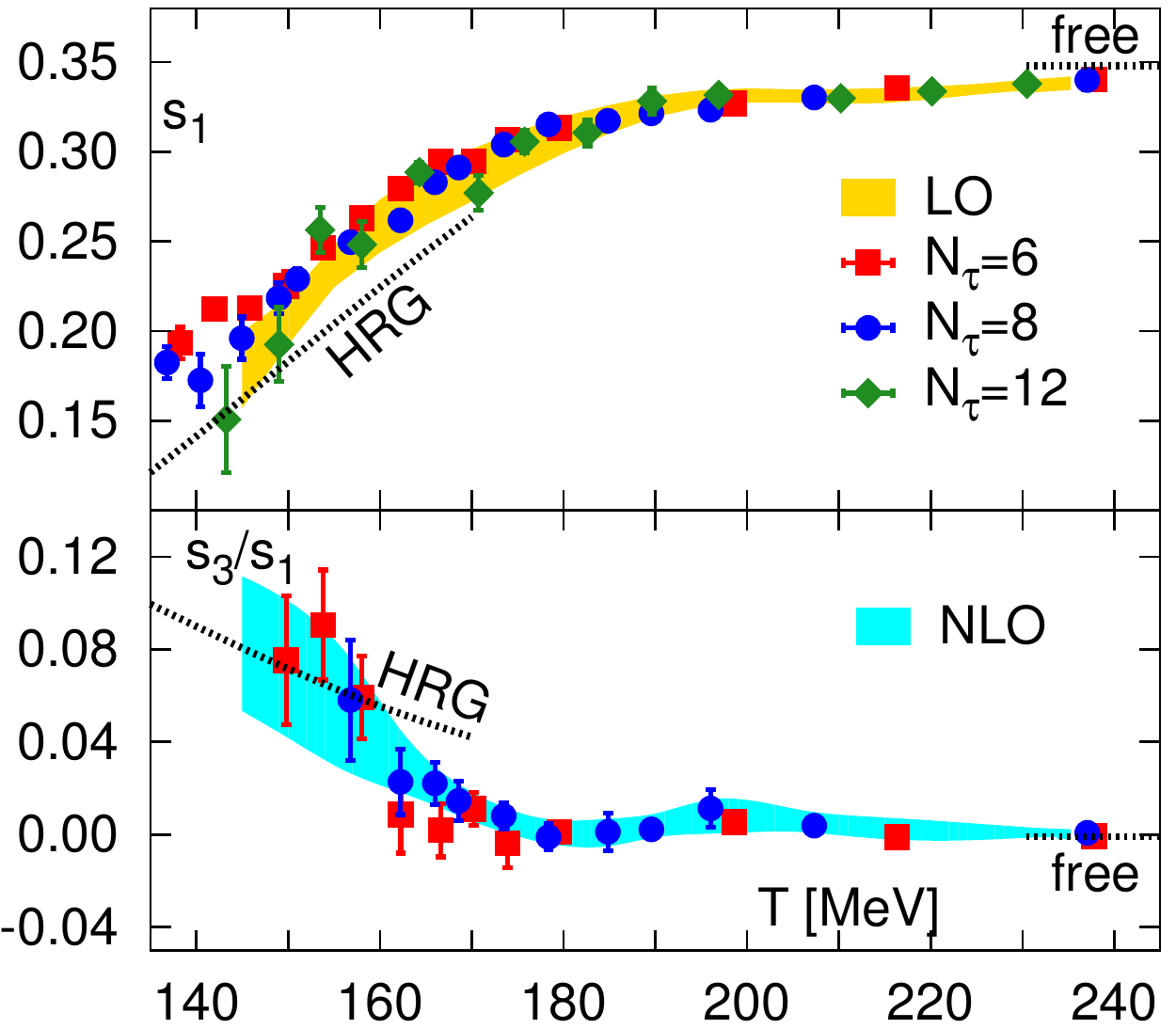} }
\subfigure[]{ \label{fig:muQS} 
\includegraphics[height=0.25\textheight,width=0.31\textwidth]{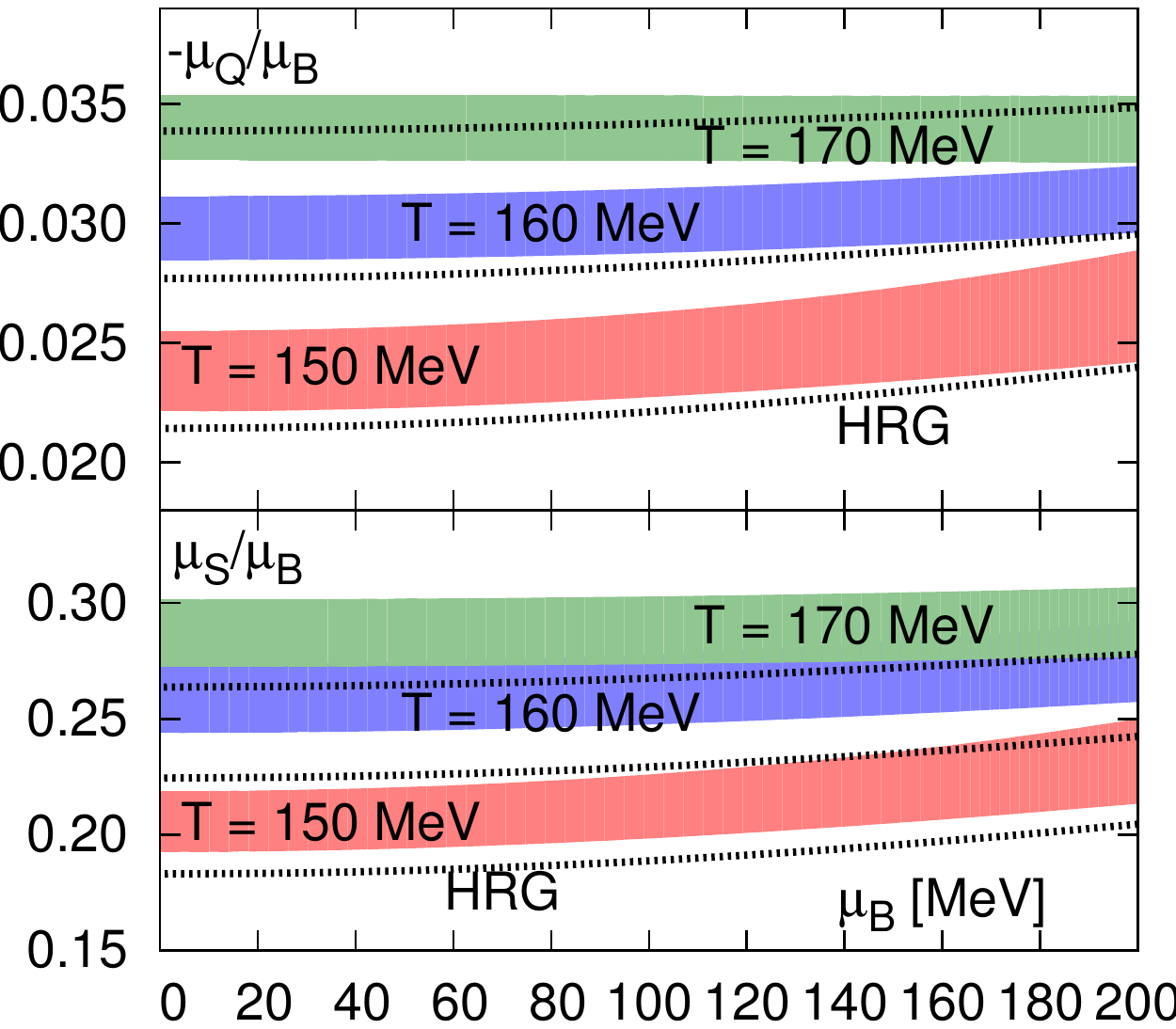} }
\caption{(a) LQCD results \cite{freeze} for the LO (top) and the  NLO (bottom) in
$\mu_B$ contributions for the electric charge chemical potential as a function of
temperature.  (b) Same as the previous panel, but for the strangeness chemical
potential. (c) Electric charge (top) and strangeness (bottom) chemical potential as a
function of $\mu_B$ for the relevant temperature range $T=150-170$ MeV.}
\end{figure}

In Heavy-Ion Collisions (HIC) experiments the measured hadrons come from the
freeze-out stage of the fireball evolution. The success of statistical hadronization
models \cite{pbm} in fitting the experimentally measured hadron yields suggests that
freeze-out conditions in HIC can be described by equilibrium thermodynamics
characterized by freeze-out temperatures ($T^f$) and chemical potentials
($\mu_B^f,\mu_Q^f,\mu_S^f$). Thus, if at all, the thermal conditions probed in HIC
corresponds to these freeze-out parameters. To capture any signature of criticality
in HIC the freeze-out must occur close to the QCD crossover/transition in the
$T-\mu_B$ plane. By now we have quite reliable knowledge regarding the location of
the chiral and deconfinement crossover of QCD in the $T-\mu_B$ plane, for moderately
small values of $\mu_B$, from first-principle LQCD calculations
\cite{strange,Tc,curv}. On the other hand, so far the freeze-out conditions of HIC
have not been determined on an equal footing but only by using model fits \cite{pbm}.
Here we introduce a new methodology for a model independent extraction of the
freeze-out parameters through a comparison between experimentally measured cumulants
of conserved charge fluctuations and LQCD calculations \cite{freeze}.

For a consistent determination of $\mu_Q^f$ and $\mu_S^f$ it necessary to realize
that these two parameters are not independent of $T^f$ and $\mu_B^f$ owing to the
initial strangeness neutrality and initial iso-spin asymmetry of the colliding nuclei
of HIC. As the net electric charge and net strangeness remain conserved throughout
the evolution of the fireball, assuming spatial homogeneity, the initial strangeness
neutrality leads to $\lang n_S \rang=0$ and the initial iso-spin asymmetry of the
colliding nuclei translates into the relation $\lang n_Q \rang=r\lang n_B \rang$.
Here, $n_X$ denotes the density of the corresponding net conserved charge $X$ and
$r=N_p/(N_p+N_n)$ is the ratio of the total number of protons to the total number of
protons and neutrons of the initially colliding nuclei. For the RHIC Au-Au and the
LHC Pb-Pb collisions $r=0.4$ provides a good approximation and will be used in our
consistent determination for $\mu_Q^f$ and $\mu_S^f$. Through a Taylor series
expansion of $\lang n_X \rang$ in powers of $(\mu_B,\mu_Q,\mu_S)$ up to
$\mathcal{O}(\mu_X^3)$ and by imposing the above constraints it is possible to write
down $\mu_Q$ and $\mu_S$ in terms of the $T^f$ and $\mu_B^f$ \cite{freeze}
\beq
\mu_Q(T,\mu_B) = q_1(T)\mu_B + q_3(T)\mu_B^3 + \mathcal{O}(\mu_B^5)
\;, \quad 
\mu_S(T,\mu_B) = s_1(T)\mu_B + s_3(T)\mu_B^3 + \mathcal{O}(\mu_B^5)
\;. \label{eq:muQ-muS}
\eeq
In \fig{fig:qi} and \fig{fig:si} we show LQCD results for the Leading Order (LO)
contribution $q_1(T)$ and $s_1(T)$ (top panel) and the Next-to-Leading Order (NLO)
contribution $q_3(T)$ and $s_3(T)$ (bottom panel) to $\mu_Q$ and $\mu_S$,
respectively. The NLO contributions are below $10\%$ and are well controlled for a
baryon chemical potential $\mu_B\lesssim200$ MeV, \ie\ for RHIC energies down to
$\sqrt{s_{NN}}\gtrsim19.6$ GeV. The complete LO plus NLO results for $\mu_Q(T,\mu_B)$
(top panel) and $\mu_S(T,\mu_B)$ (bottom panel) as a function of $\mu_B$ for the
relevant temperature range $T=150-170$ MeV are shown in \fig{fig:muQS}. Note that
around $T\approx157$ MeV the LQCD results for $\mu_S/\mu_B\approx0.24$ is quite close
to that extracted from the statistical model based fits of the strange baryons to
anti-baryons ratios measures by the STAR experiment as part of the RHIC BES program
\cite{zhao}. This observation not only confirms that strangeness neutrality is also
realized during these HIC but also provides a hint for the value of the freeze-out
temperature. 

\begin{figure}[t!]
\subfigure[]{ \label{fig:R31Q} 
\includegraphics[height=0.25\textheight,width=0.48\textwidth]{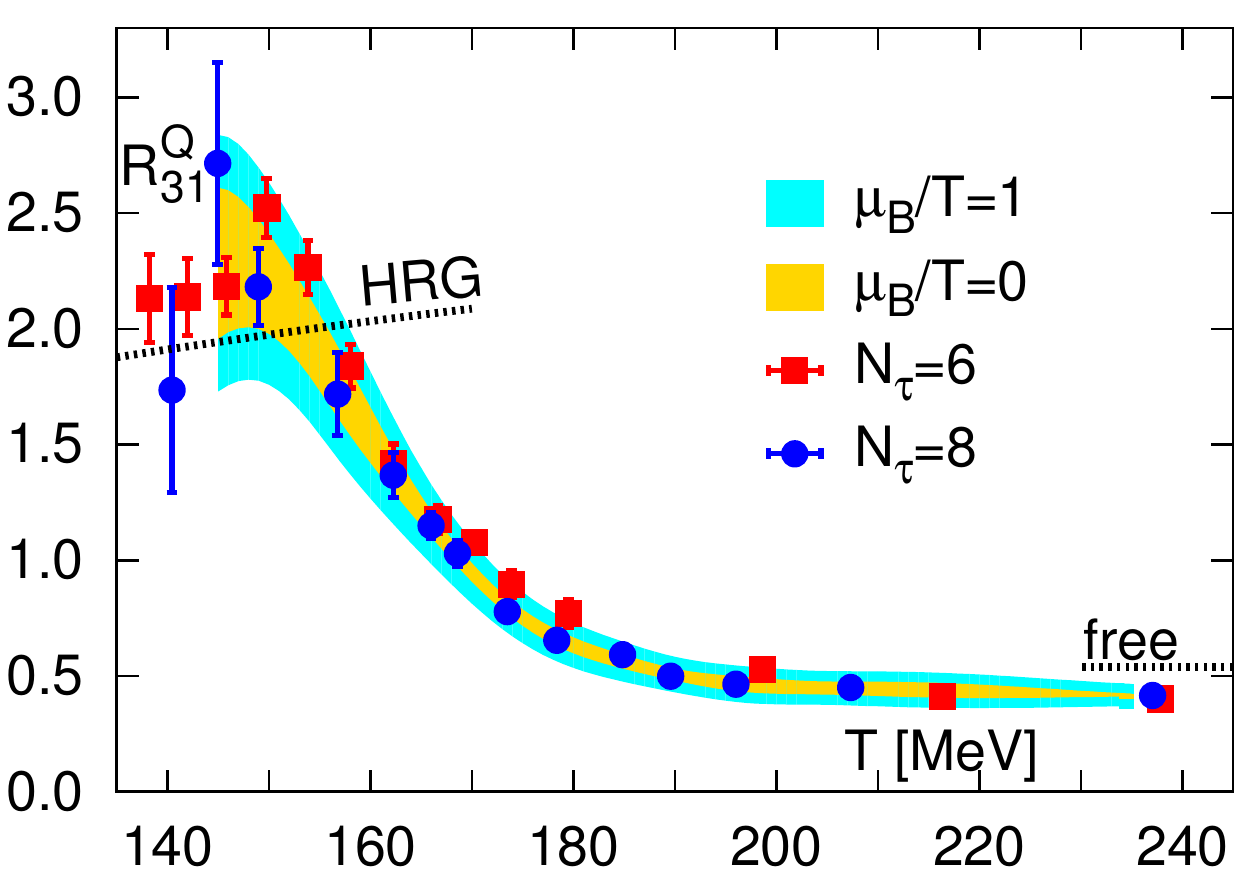} }
\subfigure[]{ \label{fig:R12Q} 
\includegraphics[height=0.25\textheight,width=0.48\textwidth]{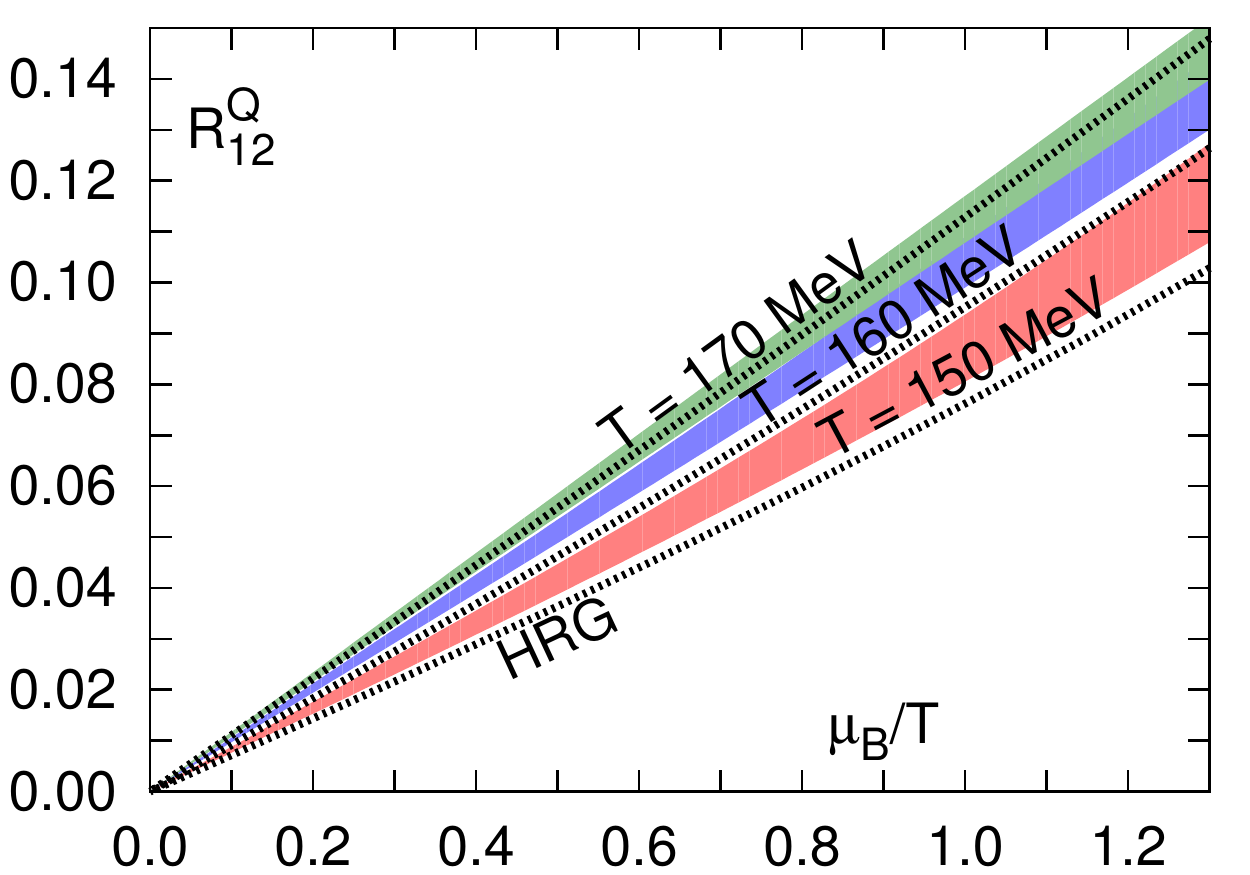} }
\caption{LQCD results \cite{freeze} for the \emph{thermometer} $R_{31}^Q$ (a) and the
\emph{baryometer} $R_{12}^Q$ (b) up to order $\mu_B^2$.}
\end{figure}

\begin{figure}[t!]
\subfigure[]{ \label{fig:R31Q-star-27} 
\includegraphics[height=0.25\textheight,width=0.48\textwidth]{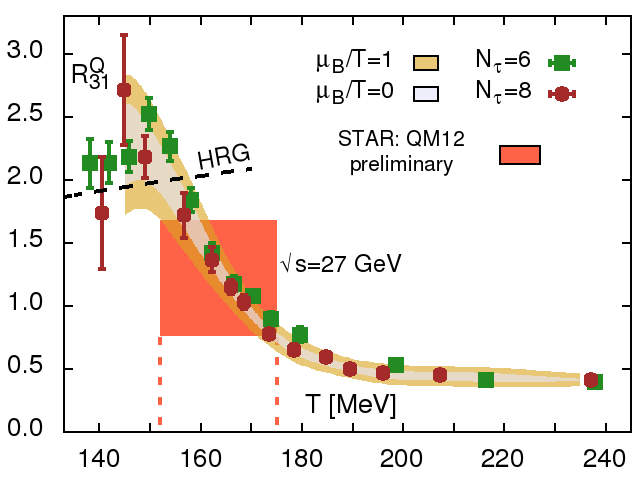} }
\subfigure[]{ \label{fig:R31Q-star-39} 
\includegraphics[height=0.25\textheight,width=0.48\textwidth]{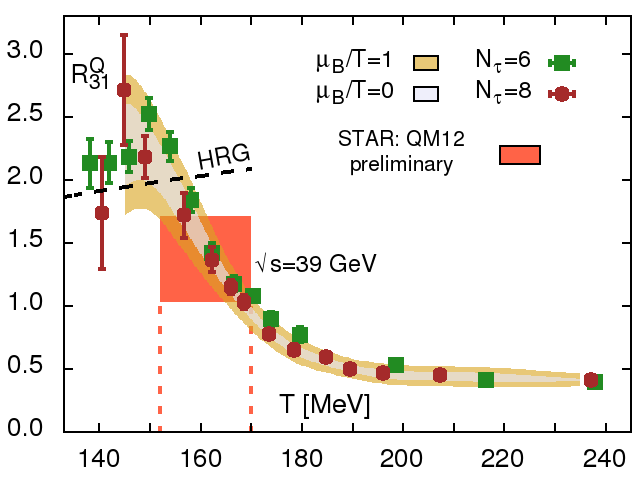} }
\caption{Comparisons between the LQCD results \cite{freeze} for the
\emph{thermometer} $R_{31}^Q$ and the ratio $(S_Q \sigma_Q^3)/M_Q$ of the cumulants
of the net electric charge fluctuation measured by the STAR experiment
\cite{star-charge} at RHIC beam energies of $\sqrt{s_{NN}}=27$ GeV (a) and
$\sqrt{s_{NN}}=39$ GeV. The overlap regions of the experimental results with the LQCD
calculations provide estimates for the freeze-out temperatures at these energies.} 
\end{figure}

To eliminate the explicit (unknown) volume factors we choose to work with the ratios
of cumulants of conserved charge fluctuations. As discussed in the Introduction, the
experimentally measurable ratios of cumulants are related to the ratios of
generalized susceptibilities. With the knowledge of $\mu_Q(T,\mu_B)$ and
$\mu_S(T,\mu_B)$ all these susceptibilities can be calculated as function of
$(T,\mu_B)$ using LQCD with a Taylor series expansion in $\mu_B$. Since the
fluctuations of the net electric can be measured both in experiments and LQCD, as an
explicit example we consider the following ratios of the cumulants of the net charge
fluctuations
\beqa
R_{31}^Q &\equiv& \frac{\chi_3^Q(T,\mu_B)}{\chi_1^Q(T,\mu_B)} =
\frac{S_Q\sigma_Q^3}{M_Q} = R_{31}^{Q,0} + R_{31}^{Q,2} \mu_B^2 + \mathcal{O}(\mu_B^4)
\label{eq:R31Q} \\
R_{12}^Q &\equiv& \frac{\chi_1^Q(T,\mu_B)}{\chi_2^Q(T,\mu_B)} = \frac{M_Q}{\sigma_Q^2} 
= R_{12}^{Q,1} \mu_B + R_{12}^{Q,3} \mu_B^3 + \mathcal{O}(\mu_B^5)
\label{eq:R12Q} \;.
\eeqa
In LO $R_{31}^Q$ is independent of $\mu_B$ while the LO term for $R_{12}^Q$ is
proportional to $\mu_B$. This suggests the use of $R_{31}^Q$ as \emph{thermometer} to
determine $T^f$ and of $R_{12}^Q$ as \emph{baryometer} to 'measure' $\mu_B^f$. In
\fig{fig:R31Q} and \fig{fig:R12Q} we show the LQCD results \cite{freeze} for the
ratio $R_{31}^Q$ and $R_{12}^Q$, respectively. In the temperature range of interest
$T=150-170$ MeV, the estimated NLO corrections for these ratios are  below 10\% and
hence these results are well under control for $\mu_B\lesssim200$ MeV.  Thus, these
LQCD data for the \emph{thermometer} $R_{31}^Q$ and the \emph{baryometer} $R_{12}^Q$
can be directly compared with corresponding experimentally measured ratios of net
charge cumulants to extract  $T^f$ and $\mu_B^f$  for RHIC energies down to
$\sqrt{s_{NN}}\gtrsim19.6$ GeV.

\begin{figure}[t!]
\subfigure[]{ \label{fig:R31Q-star-avg} 
\includegraphics[height=0.25\textheight,width=0.48\textwidth]{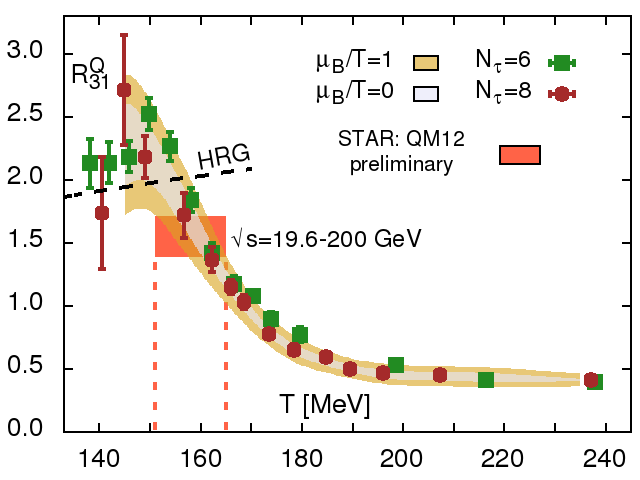} }
\subfigure[]{\label{fig:R12Q-phenix} 
\includegraphics[height=0.25\textheight,width=0.48\textwidth]{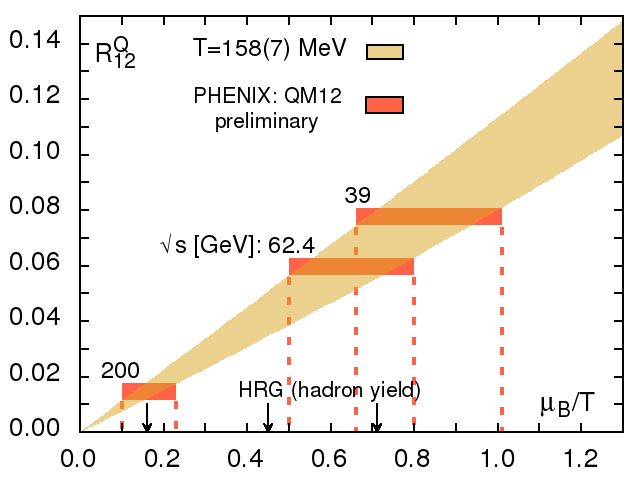} }
\caption{(a) Comparisons between the LQCD results \cite{freeze} for the
\emph{thermometer} $R_{31}^Q$ and the preliminary STAR data \cite{star-charge} for
ratio $(S_Q \sigma_Q^3)/M_Q$ of the cumulants of the net electric charge fluctuation,
averaged over the energy range $\sqrt{s_{NN}}=19.6-200$ GeV. The overlap of the
experimental results with the LQCD calculations provides an estimate for the average
freeze-out temperature $T^f=158(7)$ MeV over $\sqrt{s}=19.6-200$ GeV.  (b) LQCD
results \cite{freeze} for the \emph{baryometer} $R_{12}^Q$ as a function of $\mu_B/T$
compared with the preliminary PHENIX data \cite{phenix-charge} for $M_Q/\sigma_Q^2$
in the temperature range $T^f=158(7)$ MeV. The overlap regions of the experimentally
measured results with the LQCD calculations provide estimates for the freeze-out
chemical potential $\mu_B^f$ for a given $\sqrt{s_{NN}}$. The arrows indicate the
values of $\mu_B^f/T^f$ obtained from traditional statistical model fits to
experimentally measured hadron yields \cite{Cleymans:2005xv}.}
\end{figure}

As practical demonstrations of this methodology, in \fig{fig:R31Q-star-27} and
\fig{fig:R31Q-star-39} we show comparisons of the LQCD results for the
\emph{thermometer} $R_{31}^Q$ with the preliminary STAR data \cite{star-charge} for
the corresponding ratio $(S_Q \sigma_Q^3)/M_Q$ of the cumulants of the net charge
fluctuation for the RHIC beam energies of $\sqrt{s_{NN}}=27$ GeV and
$\sqrt{s_{NN}}=39$ GeV, respectively. The freeze-out temperature $T^f$ for a given
beam energy can be extracted from the temperature range over which the LQCD
calculations and the experimental data overlap. It is clear that the uncertainties of
the preliminary experimental data are too large to extract the $\sqrt{s_{NN}}$
dependence of $T^f$. Thus, for the illustration of the determination of the $\mu_B^f$
we use the preliminary STAR data for $(S_Q \sigma_Q^3)/M_Q$ averaged over the beam
energy range $\sqrt{s_{NN}}=19.6-200$ GeV and compare that with the LQCD results of
$R_{31}^Q$ in \fig{fig:R31Q-star-avg}. In this way we can determine an average
freeze-out temperature of $T^f=158(7)$ MeV for the RHIC beam energies of
$\sqrt{s_{NN}}=19.6-200$ GeV. In this energy range also the traditional statistical
model fits \cite{pbm,Cleymans:2005xv}  yield an almost constant value of $T^f$ .
Hence, for the illustrative purpose our use of an average $T^f$ is quite justified.
Furthermore, we only use RHIC data for energies down to $\sqrt{s_{NN}}=19.6$ MeV as
for smaller energies $\mu_B^f$ becomes too large to justify the use of our LQCD
calculations which are performed only up to NLO in $\mu_B$.

\begin{figure}[t!]
\subfigure[]{\label{fig:R12Q-star} 
\includegraphics[height=0.25\textheight,width=0.48\textwidth]{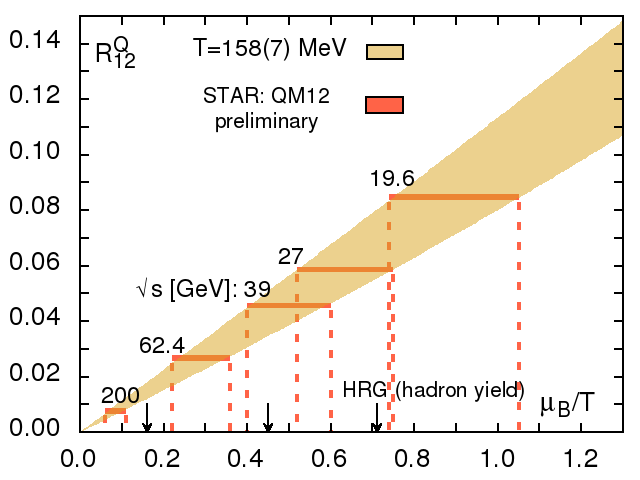} }
\subfigure[]{\label{fig:R12B-star} 
\includegraphics[height=0.25\textheight,width=0.48\textwidth]{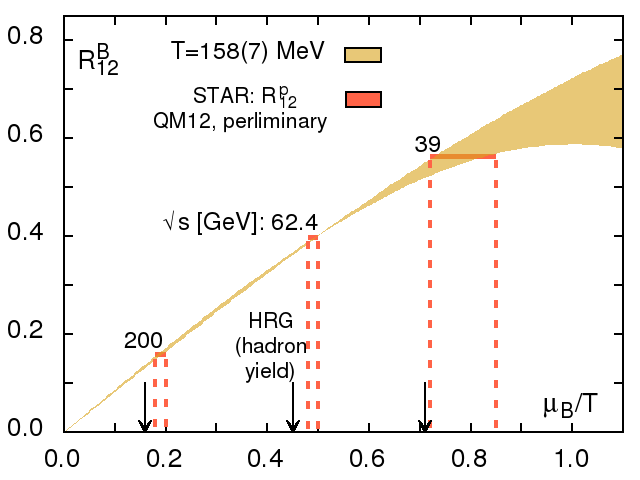} }
\caption{(a) Comparison of the LQCD results \cite{freeze} for the \emph{baryometer}
$R_{12}^Q$, in the temperature range $T^f=158(7)$ MeV, with the preliminary STAR data
\cite{star-charge} for the ratio $M_Q/\sigma_Q^2$ of the cumulants of net charge
fluctuation at several RHIC beam energies $\sqrt{s_{NN}}$.  (b) LQCD results
\cite{freeze} for the $R_{12}^B$, in the temperature range $T^f=158(7)$ MeV, as a
function of $\mu_B/T$ compared with the preliminary STAR data \cite{star-proton} for
the ratio $M_p/\sigma_p^2$ of the cumulants of net proton fluctuation.  The overlap
regions of the experimentally measured and LQCD results provide estimates for the
freeze-out chemical potential $\mu_B^f$ for a given $\sqrt{s_{NN}}$. The arrows
indicate the values of $\mu_B^f/T^f$ obtained from traditional statistical model fits
to experimentally measured hadron yields \cite{Cleymans:2005xv}.}
\label{fig:R12-star}
\end{figure}

In \fig{fig:R12Q-phenix} we show the LQCD results for the \emph{baryometer}
$R_{12}^Q$ as a function of $\mu_B/T$ in the previously determined average freeze-out
temperature range of $T^f=158(7)$ and compare it to the ratio $M_Q/\sigma_Q^2$ of the
cumulants of the net charge fluctuation measured by the PHENIX
collaboration~\cite{phenix-charge} at several RHIC beam energies. Similar comparisons
with the preliminary STAR data~\cite{star-charge} for $M_Q/\sigma_Q^2$ are shown in
\fig{fig:R12Q-star}. The freeze-out baryon chemical potential $\mu_B^f/T^f$ can be
determined from the overlap region of the LQCD and experimental results. Thus, by
applying this methodology in a similar manner for each beam energy the corresponding
freeze-out temperature and baryon chemical potential can be obtained in a completely
model independent way through direct comparisons of the LQCD and HIC experiments.

\begin{figure}[t!]
\begin{center}
\includegraphics[height=0.25\textheight,width=0.9\textwidth]{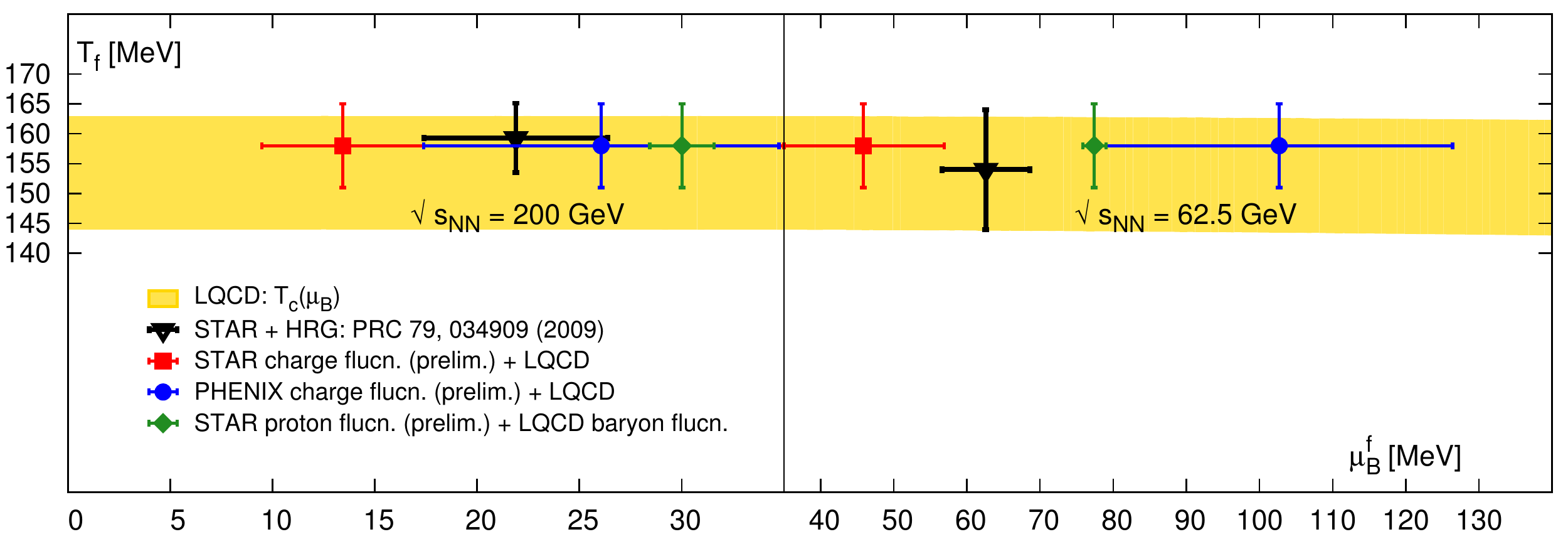}
\end{center}
\caption{Freeze-out temperatures $T^f$ and baryon chemical potentials $\mu_B^f$
obtained through direct comparisons between LQCD calculations and the preliminary
STAR and PHENIX data for cumulants of net charge and net proton fluctuations. The
shaded region indicate the LQCD results \cite{strange,Tc,curv} for the
chiral/deconfinement temperature $T_c$ as a function of the baryon chemical
potential.} 
\label{fig:tc-fo} 
\end{figure}

If the freeze-out stage of HIC is indeed described by equilibrium thermodynamics then
thermodynamic consistency demands that $T^f$ and $\mu_B^f$ determined through
different observables should produce the same values for these thermodynamic
parameters. For example, instead of the net electric charge one may use the
fluctuation of net baryon number and use LQCD results for
$R_{12}^B=\chi_1^B/\chi_2^B$ as the \emph{baryometer}. Despite the caveat that
cumulants of net proton fluctuation may be quantitatively different from the
cumulants of net baryon number fluctuation \cite{Bzdak:2012ab,Kitazawa}, in
\fig{fig:R12B-star} we present a comparison between the LQCD results for  $R_{12}^B$
and the preliminary STAR data \cite{star-proton} for the ratio $M_p/\sigma_p^2$ of
the cumulants of net proton fluctuation. Unfortunately, as can be seen from
\fig{fig:R12Q-phenix} and \fig{fig:R12-star}, the freeze-out baryon chemical
potential obtained from all these experimental measurements are not consistent with
each other at present. Furthermore, they are also not consistent with the freeze-out
baryon chemical potential obtained obtained form the traditional statistical model
fits to the experimentally measured hadron yields \cite{Cleymans:2005xv}. To
illustrate this more clearly in \fig{fig:tc-fo} we show  the freeze-out parameters
$T^f$ and $\mu_B^f$ extracted by comparing LQCD calculations with the preliminary
STAR and PHENIX results for the cumulants of net charge fluctuations as well as with
the preliminary STAR data for the cumulants of net proton fluctuations. While these
results differ from each other and from that obtained using the statistical model
fits to the experimentally measured hadron yields \cite{star-fo}, it is tantalizing
to see that all these results lie within the chiral/deconfinement crossover region,
$T_c(\mu_B)=\lrnd154(9)-[0.0066(7)/154(9)]\mu_B^2\rrnd$ MeV, obtained from LQCD
calculations \cite{strange,Tc,curv}. This makes us hopeful that the HIC collision
experiments may signal presence of criticality in the QCD phase diagram in the
$T-\mu_B$ plane.  

While such direct comparisons between the LQCD calculations and HIC experiments may
open up many new opportunities, at present, one has to be somewhat cautious. The LQCD
calculations of generalized susceptibilities are performed using a grand-canonical
ensemble approach in the thermodynamic limit. It is a-priori not evident that this is
also applicable to conditions met in a heavy ion collision. Thus while comparing our
results with experimental ones we must make sure that effects of conservation laws
due to finite system sizes, acceptance cuts \cite{Bzdak:2012ab,Bzdak:2012an} \etc\ do
not invalidate the grand canonical ensemble approach. These questions are currently
being addressed in experimental analysis \cite{star-charge,phenix-charge,star-proton}
and hopefully will be resolved soon.

%------------------------------------------------------------------------------------
%       references
%------------------------------------------------------------------------------------

%------------------------------------------------------------------------------------
%       end
%------------------------------------------------------------------------------------

\begin{thebibliography}{99}

\bibitem{strange} 
A. Bazavov {\it et al.}, arXiv:1304.7220 [hep-lat].

\bibitem{freeze} 
A. Bazavov {\it et al.}, Phys.\ Rev.\ Lett.\  {\bf 109}, 192302 (2012).

\bibitem{hrg}
A. Bazavov {\it et al.}, Phys.\ Rev.\ D {\bf 86}, 034509 (2012).

\bibitem{Tc}
A. Bazavov {\it et al.}, Phys.\ Rev.\ D {\bf 85}, 054503 (2012).

\bibitem{pbm}
For a review see: P. Braun-Munzinger, K. Redlich, and J. Stachel, In {\it Hwa, R.C.
(ed.) {\it et al.}: Quark gluon plasma} 491-599, [nucl-th/0304013].

\bibitem{Vuorinen}
J. O. Andersen, S. Mogliacci, N. Su and A. Vuorinen, Phys.\ Rev.\ D {\bf 87}, 074003
(2013).

\bibitem{curv}
O. Kaczmarek {\it et al.}, Phys.\ Rev.\ D {\bf 83}, 014504 (2011).

\bibitem{zhao}
F. Zhao, this proceedings.

\bibitem{star-charge}
D. McDonald [STAR Collaboration], Nucl.\ Phys.\ A904-905 {\bf 2013}, 907c (2013).

\bibitem{phenix-charge}
J. T. Mitchell [PHENIX Collaboration], Nucl.\ Phys.\ A904-905 {\bf 2013}, 903c
(2013).

\bibitem{Bzdak:2012ab} 
A. Bzdak and V. Koch, Phys.\ Rev.\ C {\bf 86}, 044904 (2012).

\bibitem{Kitazawa} 
M. Kitazawa and M. Asakawa, Phys.\ Rev.\ C {\bf 85}, 021901 (2012); Phys.\ Rev.\ C
{\bf 86}, 024904 (2012) [Erratum-ibid.\ C {\bf 86}, 069902 (2012)]. 

\bibitem{star-proton} 
X. Luo [STAR Collaboration], Nucl.\ Phys.\ A904-905 {\bf 2013}, 911c (2013).

\bibitem{Cleymans:2005xv} 
J. Cleymans, H. Oeschler, K. Redlich and S. Wheaton, Phys.\ Rev.\ C {\bf 73}, 034905
(2006).

\bibitem{star-fo}
B. I. Abelev {\it et al.}  [STAR Collaboration], Phys.\ Rev.\ C {\bf 79}, 034909
(2009)

\bibitem{Bzdak:2012an} 
A. Bzdak, V. Koch and V. Skokov, Phys.\ Rev.\ C {\bf 87}, 014901 (2013).

\end{thebibliography}
\end{document}